\documentclass[amsmath,showpacs,nofootinbib,11pt,superscriptaddress]{revtex4-2}
\usepackage{graphicx}
\usepackage{dcolumn}
\usepackage{bm}
\usepackage{subcaption}
\usepackage{color} 
\usepackage{slashed}
\usepackage{float}
\usepackage{amsfonts}
\usepackage{multirow}
\begin{document}
\newcommand{\overlr}{\stackrel{\leftrightarrow}{\partial}}
\newcommand{\hs}{\hspace*{0.5cm}}
\newcommand{\vs}{\vspace*{0.5cm}}
\newcommand{\be}{\begin{equation}}
\newcommand{\ee}{\end{equation}}
\newcommand{\bea}{\begin{eqnarray}}
\newcommand{\eea}{\end{eqnarray}}
\newcommand{\ben}{\begin{enumerate}}
\newcommand{\een}{\end{enumerate}}
\newcommand{\bde}{\begin{widetext}}
\newcommand{\ede}{\end{widetext}}
\newcommand{\nn}{\nonumber}
\newcommand{\crn}{\nonumber \\}
\newcommand{\Tr}{\mathrm{Tr}}
\newcommand{\non}{\nonumber}
\newcommand{\noi}{\noindent}
\newcommand{\al}{\alpha}
\newcommand{\la}{\lambda}
\newcommand{\bet}{\beta}
\newcommand{\ga}{\gamma}
\newcommand{\va}{\varphi}
\newcommand{\om}{\omega}
\newcommand{\pa}{\partial}
\newcommand{\+}{\dagger}
\newcommand{\fr}{\frac}
\newcommand{\bc}{\begin{center}}
\newcommand{\ec}{\end{center}}
\newcommand{\Ga}{\Gamma}
\newcommand{\de}{\delta}
\newcommand{\De}{\Delta}
\newcommand{\ep}{\epsilon}
\newcommand{\varep}{\varepsilon}
\newcommand{\ka}{\kappa}
\newcommand{\La}{\Lambda}
\newcommand{\si}{\sigma}
\newcommand{\Si}{\Sigma}
\newcommand{\ta}{\tau}
\newcommand{\up}{\upsilon}
\newcommand{\Up}{\Upsilon}
\newcommand{\ze}{\zeta}
\newcommand{\ps}{\psi}
\newcommand{\Ps}{\Psi}
\newcommand{\ph}{\phi}
\newcommand{\vph}{\varphi}
\newcommand{\Ph}{\Phi}
\newcommand{\Om}{\Omega}
\newcommand{\AdrHEPC}{Phenikaa Institute for Advanced Study, Phenikaa University, Duong Noi, Hanoi 100000, Vietnam}
\newcommand{\AdrH}{Institute of Physics, Vietnam Academy of Science and Technology,
10 Dao Tan, Giang Vo, Hanoi 100000, Vietnam}

\title{Two-component dark matter from a flavor-dependent $U(1)$ gauge extension}

\author{N. T. Duy}
\email{ntduy@iop.vast.vn}
\affiliation{\AdrH}
\author{Duy H. Nguyen}
\email{duy.nguyenhoang@phenikaa-uni.edu.vn}
\affiliation{\AdrHEPC}
\author{Do Thi Ha}
\email{hak37sp2spl@gmail.com}
\affiliation{Tran Quoc Tuan University - First Army Academy, Doai Phuong, Hanoi 100000, Vietnam}
\author{Duong Van Loi}
\email{loi.duongvan@phenikaa-uni.edu.vn (corresponding author)}
\affiliation{\AdrHEPC} 

\date{\today}

\begin{abstract}
We revisit the dark matter phenomenology of a flavor-dependent $U(1)_X$ gauge extension of the Standard Model, where anomaly cancellation predicts the existence of exactly three fermion generations and requires the presence of three right-handed neutrinos. In Ref.~\cite{VanLoi:2023utt}, a strong hierarchy between the vacuum expectation values of two singlet scalars, $\La_2 \gg \La_1$, renders all $\mathbb{Z}_2$-odd scalar states heavy, resulting in a two-component dark matter scenario composed exclusively of fermions. In the present work, we relax this simplifying assumption and consider a more general mass spectrum. In particular, scalar mixing can naturally lead to a situation in which the lightest $\mathbb{Z}_2$-odd particle is a scalar rather than a fermion. As a consequence, the model admits a qualitatively new realization of two-component dark matter consisting of one fermionic and one scalar component, in addition to the purely fermionic scenario studied previously. We perform a dedicated phenomenological analysis of these two-component dark matter realizations, focusing on the coupled thermal freeze-out dynamics and the resulting relic abundance. Constraints from the observed relic density and current direct-detection limits are taken into account, and viable regions of parameter space are identified.
\end{abstract}

\maketitle

\section{Introduction}
The existence of dark matter (DM)~\cite{ParticleDataGroup:2024cfk,Planck:2018vyg} provides compelling evidence for physics beyond the Standard Model (SM). Among various well-motivated possibilities, weakly interacting massive particles (WIMPs) remain an attractive class of candidates, as they can be probed by cosmological observations, direct- and indirect-detection experiments, as well as collider searches \cite{Jungman:1995df,Bertone:2004pz}. In recent years, scenarios involving more than one DM component have received increasing attention, since they naturally arise in many extensions of the SM and can lead to rich and distinctive phenomenological features that differ significantly from those of single-component DM~\cite{Boehm:2003ha,Zurek:2008qg,Aoki:2013gzs,Kong:2014mia,Agashe:2014yua,Kim:2016zjx,Arcadi:2016kmk,Tulin:2017ara,Biswas:2019ygr,Borah:2019epq,Nam:2020twn}.

Gauge extensions of the SM based on an additional $U(1)$ symmetry provide a simple and theoretically appealing framework for addressing several open questions, including neutrino masses and DM~\cite{VanLoi:2020kdk,VanDong:2021hhg,VanDong:2020bkg}. In particular, flavor-dependent $U(1)$ symmetries are well motivated, as there is no fundamental reason for new gauge interactions to be family universal~\cite{VanDong:2022cin,VanLoi:2023pkt,VanLoi:2023kgl,VanLoi:2025fmy}. When combined with anomaly cancellation conditions, such constructions can become highly predictive, severely  constraining the fermion content and charge assignments of the theory.

Recently, an interesting flavor-dependent $U(1)_X$ extension of the SM has been proposed, where anomaly cancellation uniquely predicts the existence of three fermion generations and necessitates the introduction of three right-handed neutrinos~\cite{VanLoi:2023utt}. After spontaneous symmetry breaking, a residual discrete symmetry $\mathbb{Z}_2$ remains, ensuring the stability of the lightest $\mathbb{Z}_2$-odd particle as a DM candidate. In addition, one $\mathbb{Z}_2$-even right-handed neutrino is accidentally stabilized as a consequence of gauge invariance. The model therefore naturally accommodates a two-component DM scenario, together with radiative neutrino mass generation via the scotogenic mechanism. In Ref.~\cite{VanLoi:2023utt}, the DM phenomenology of this framework was investigated under the simplifying assumption of a strong hierarchy between the vacuum expectation values (VEVs) of two singlet scalars, $\Lambda_2 \gg \Lambda_1$. In this limit, all $\mathbb{Z}_2$-odd scalar states acquire large masses proportional to $\Lambda_2$, whereas the $\mathbb{Z}_2$-odd right-handed neutrinos remain much lighter, with masses controlled by $\Lambda_1$. As a result, the DM sector consists exclusively of two fermionic components, and the corresponding two-component DM phenomenology was analyzed in detail.

In the present work, we relax the simplifying assumption adopted in Ref.~\cite{VanLoi:2023utt} and consider a more general mass spectrum. In particular, we allow for the possibility that the lightest $\mathbb{Z}_2$-odd particle is a scalar rather than a fermion. This extension qualitatively enlarges the DM sector of the model. As a consequence, in addition to the purely fermionic two-component DM scenario studied previously, the model admits a new realization of two-component DM composed of one fermionic and one scalar component. The presence of a scalar DM particle opens up new annihilation and coannihilation channels and can significantly affect the coupled thermal evolution of the two DM components. The purpose of this paper is to investigate the phenomenology of two-component DM in this extended setup. We analyze both the fermion–fermion and fermion–scalar realizations, focusing on their coupled thermal freeze-out dynamics and the resulting relic abundance. The viable parameter space is further confronted with current direct-detection constraints and cosmological measurements of the DM relic density.

This paper is organized as follows. In Sec.~\ref{sec:model}, we briefly review the model and discuss the general scalar mass spectrum, including mixing effects, together with the gauge sector. Section~\ref{sec:DM} discusses the residual discrete symmetry and classifies the possible two-component DM scenarios. In Sec.~\ref{sec:relic}, we investigate the relic abundance and the coupled freeze-out dynamics, and confront the viable parameter space with current experimental constraints. Finally, we summarize our results and conclude in Sec.~\ref{sec:conclusion}.

\section{\label{sec:model}The model}
\subsection{Gauge anomaly cancellation and particle content}
As proposed in Ref.~\cite{VanLoi:2023utt}, we extend the SM gauge symmetry by an additional Abelian factor, $U(1)_X$, where the $X$ charge is taken to be a linear combination of the usual baryon ($B$) and lepton ($L$) numbers, with family-dependent coefficients. Explicitly, the $U(1)_X$ charge is defined as $X_i=x_iB+y_iL$, where $i$ denotes the fermion family index. Further, we restrict $x_{1,2,\cdots,n}=-3z$, $x_{n+1,n+2,\cdots,n+m=N_f}=3z$, and $y_{1,2,\cdots,N_f}=z$, where $n$ and $m$ are  integers, and $z$ is an arbitrary nonzero parameter. 

The cancellation of the $[SU(2)_L]^2U(1)_X$ gauge anomaly requires $N_f=3(n-m)$, implying that the number of fermion families must be a multiple of the color number, $3$. Combined with the requirement of QCD asymptotic freedom, $N_f\leq 8$, only two solutions are allowed, namely $N_f=3$ and $N_f=6$. We choose $N_f=3$, consistent with experimental observations, which corresponds to $n=2$ and $m=1$. 

Furthermore, the cancellation of the mixed gravitational anomaly $[\mathrm{Gravity}]^2U(1)_X$ and the cubic gauge anomaly $[U(1)_X]^3$ requires the inclusion of three right-handed neutrinos, $\nu_{1,2,3R}$, as fundamental fields of the model. Their $U(1)_X$ charge assignments are given by $X(\nu_{1,2R})=4z$ and $X(\nu_{3R})=-5z$.\footnote{A simpler choice, $X(\nu_{1,2,3R}) = z$, would lead to a more minimal scalar sector, but would not directly provide viable DM candidates.} The complete fermion content and their corresponding $U(1)_X$ charges are summarized in Table~\ref{fermion}.
\begin{table}[h]
\bc
\begin{tabular}{l|cccccccccccc}
\hline\hline
Field &$l_{aL}=\begin{pmatrix}
\nu_{aL}\\e_{aL}
\end{pmatrix}$ & $e_{aR}$ & $q_{\al L}=\begin{pmatrix} u_{\al L}\\ d_{\al L}\end{pmatrix}$ & $u_{\al R}$ & $d_{\al R}$ & $q_{3L}=\begin{pmatrix} u_{3L}\\ d_{3L}\end{pmatrix}$ & $u_{3R}$ & $d_{3R}$ & $\nu_{1,2R}$ & $\nu_{3R}$\\
\hline
$U(1)_X$ &$z$ & $z$ & $-z$ & $-z$ & $-z$ & $z$ & $z$ & $z$ & $4z$ & $-5z$\\
\hline\hline
\end{tabular}
\caption{\label{fermion}Fermion content and $U(1)_X$ charge assignments, where $a=1,2,3$ and $\alpha=1,2$ denote family indices.}
\ec
\end{table}

Concerning the scalar sector, in addition to the SM Higgs doublet $H$, two gauge-singlet scalars, $\chi_1$ and $\chi_2$, are introduced to generate Majorana masses for the right-handed neutrinos and to spontaneously break the $U(1)_X$ symmetry. These fields acquire nonzero VEVs,
\be \langle H\rangle=\begin{pmatrix}
0\\
\fr{v}{\sqrt2}\end{pmatrix},\hs \langle\chi_1\rangle=\fr{\La_1}{\sqrt2},\hs \langle\chi_2\rangle=\fr{\La_2}{\sqrt2},\label{vevs}\ee
with $\La_{1,2}\gg v\simeq 246~\mathrm{GeV}$, ensuring consistency with SM phenomenology. Furthermore, owing to the $X$ charge assignments, Dirac neutrino masses are naturally suppressed. To generate the observed active neutrino masses via the scotogenic mechanism, two inert scalars---a doublet $\phi$ and a singlet $\eta$---are required. The full scalar content and their quantum numbers are summarized in Table \ref{scalar}.
\begin{table}[h]
\bc
\begin{tabular}{l|cccccccc}
\hline\hline
Multiplet & $SU(3)_C\otimes SU(2)_L\otimes U(1)_Y$ & $U(1)_X$\\ \hline 
$H=(H^+,H^0)^T$ & $(1, 2,1/2)$ & $0$\\
$\chi_1$ & $(1, 1,0)$ & $-8z$\\
$\chi_2$ & $(1, 1,0)$ & $10z$\\
$\ph=(\ph^0,\ph^-)^T$ & $(1, 2,-1/2)$ & $-3z$\\
$\eta$ & $(1, 1,0)$ & $-5z$\\
\hline\hline
\end{tabular}
\caption[]{\label{scalar}Scalar content and their quantum numbers.}
\ec
\end{table}

\subsection{Particle spectrum of the scalar and gauge sectors}
With the scalar content listed in Table \ref{scalar}, the most general renormalizable scalar potential of the model is given by~\cite{VanLoi:2023utt}
\begin{align}
V =&\, \mu_0^2H^\dag H + \mu^2_1\chi_1^*\chi_1 + \mu^2_2\chi_2^*\chi_2 + \la_0(H^\dag H)^2 + \la_1(\chi_1^*\chi_1)^2 + \la_2(\chi_2^*\chi_2)^2 \crn 
&+ (H^\dag H)[\la_3(\chi_1^*\chi_1) + \la_4(\chi_2^*\chi_2)] + \la_5(\chi_1^*\chi_1)(\chi_2^*\chi_2)\crn
&+\mu^2_3\ph^\dag\ph + \mu^2_4\eta^*\eta + \la_6(\ph^\dag\ph)^2 + \la_7(\eta^*\eta)^2 + (H^\dag H)[\la_8(\ph^\dag\ph) + \la_9(\eta^*\eta)] \crn 
&+ (\chi_1^*\chi_1)[\la_{10}(\ph^\dag\ph) + \la_{11}(\eta^*\eta)]+ (\chi_2^*\chi_2)[\la_{12}(\ph^\dag\ph) + \la_{13}(\eta^*\eta)]\crn
&+ \la_{14}(\ph^\dag\ph)(\eta^*\eta) + \la_{15}(H^\dag\ph)(\ph^\dag H)+\mu(\chi_2\eta\eta+\mathrm{H.c.})+\la(H\ph\eta\chi_1^*+\mathrm{H.c.}),\label{poten}
\end{align}
where the $\la$’s denote dimensionless couplings and the $\mu$’s are parameters with mass dimension. For simplicity, all parameters are taken to be real, without loss of generality. To ensure that the scalar potential is bounded from below and that the vacuum structure follows Eq.~(\ref{vevs}), we impose $\mu^2_{0,1,2}<0$, $\mu^2_{3,4}>0$, $|\mu_0|\ll |\mu_{1,2}|$, and $\la_{0,1,2,6,7}>0$.

The scalar spectrum of the model in the limit $\La_2 \gg \La_1$ was presented in Ref.~\cite{VanLoi:2023utt}. In this limit, the model contains eight massive scalar states, namely $h$, $H_{1,2}$, $R_{1,2}$, $I_{1,2}$, and $H^\pm$, together with a physical Goldstone boson $\mathbb{A}$. In addition, three massless Goldstone bosons, $G_W$, $G_Z$, and $G_{Z'}$, are absorbed by the gauge bosons $W$, $Z$, and $Z'$, respectively. In the general case without assuming a strong hierarchy between $\La_1$ and $\La_2$, the scalar spectrum can be written as
\begin{align}
H &\simeq \begin{pmatrix}
G^+_W\\
\frac{1}{\sqrt2}(v+h+iG_Z)
\end{pmatrix},\\
\chi_1 &\simeq \frac{1}{\sqrt2}(\La_1 + c_{\epsilon_1} H_1+s_{\epsilon_1} H_2+i(c_{\epsilon_2}\mathbb{A}+s_{\epsilon_2} G_{Z'})),\\
\chi_2 &\simeq \frac{1}{\sqrt2}(\La_2-s_{\epsilon_1} H_1+c_{\epsilon_1} H_2-i(s_{\epsilon_2}\mathbb{A}-c_{\epsilon_2} G_{Z'})), \\
\ph &= \begin{pmatrix}
\frac{1}{\sqrt2}(c_{\epsilon_R} R_1+s_{\epsilon_R} R_2+i(c_{\epsilon_I} I_1+s_{\epsilon_I} I_2))\\ H^-
\end{pmatrix},\\
\eta &= -\frac{1}{\sqrt2}(s_{\epsilon_R} R_1-c_{\epsilon_R} R_2+i(s_{\epsilon_I} I_1-c_{\epsilon_I} I_2)),
\end{align}
where the mixing angles satisfy 
\begin{align}
t_{2\epsilon_1} & =\fr{\la_5\La_1\La_2}{\la_2\La_2^2-\la_1\La_1^2},\hs t_{\epsilon_2}=\fr{4\La_1}{5\La_2},\\
t_{2\epsilon_R} & =\fr{\la v\La_1}{M_1^2-M_2^2-\sqrt2\mu\La_2},\hs t_{2\epsilon_I} =\fr{\la v\La_1}{M_2^2-M_1^2-\sqrt2\mu\La_2},
\end{align}
in which $M^2_1=(2\mu^2_3+\la_{10}\La^2_1+\la_{12}\La^2_2+\la_8v^2)/2$ and $M^2_2=(2\mu^2_4+\la_{11}\La^2_1+\la_{13}\La^2_2+\la_9v^2)/2$.\footnote{Throughout this work, we adopt the shorthand notations $\sin x = s_x$, $\cos x = c_x$, and $\tan x = t_x$ for any angle $x$.} The mass of the physical scalars is approximately given by
\begin{align} m^2_h&\simeq 2\left[\la_0-\fr{\la_2\la_3^2+(\la_1\la_4-\la_3\la_5)\la_4}{4\la_1\la_2-\la_5^2}\right]v^2,\\
m^2_{H_{1,2}}&\simeq \la_1\La_1^2+\la_2\La_2^2\mp\sqrt{(\la_1\La_1^2-\la_2\La_2^2)^2+\la_5^2\La_1^2\La_2^2},\\
m^2_{R_1,I_1}&\simeq M^2_1-\fr{\la^2 v^2 \La^2_1}{4(M^2_2-M^2_1\pm\sqrt{2}\mu \La_2)}, \label{mri1}\\
m^2_{R_2,I_2}&\simeq M^2_2\pm\sqrt{2}\mu \La_2+\fr{\la^2 v^2 \La^2_1}{4(M^2_2-M^2_1\pm\sqrt{2}\mu \La_2)}, \label{mri2}\\
m^2_{H^\pm}&\simeq\mu_3^2+\frac{\la_{10}}{2}\La_1^2+\frac{\la_{12}}{2}\La_2^2 \label{mhc}.
\end{align}
Because of the condition $\La_{1,2}\gg v$, only the scalar field $h$ has a mass at the electroweak scale and can be identified with the SM Higgs boson. Also, the mixing angles $\epsilon_{R,I}$ are strongly suppressed. The physical Goldstone boson $\mathbb{A}$ is phenomenologically harmless and can participate in DM annihilation processes, as discussed in Ref.~\cite{VanLoi:2023utt}.

Concerning the gauge boson sector, the model predicts a new neutral gauge boson, denoted as $Z'$, in addition to the SM gauge bosons. Since the SM Higgs doublet $H$ does not carry the $X$ charge and the singlet scalars $\chi_{1,2}$ have zero hypercharge, there is no tree-level mass mixing between $Z'$ and the SM neutral gauge bosons. Neglecting the kinetic mixing effects between the $U(1)_Y$ and $U(1)_X$ gauge fields, the new neutral gauge boson $Z'$ is identified with the $U(1)_X$ gauge boson and acquires a mass at the new-physics scale,
\be m_{Z'}^2=4g_X^2z^2(16\La_1^2+25\La_2^2),\ee
where $g_X$ denotes the gauge coupling associated with the $U(1)_X$ symmetry~\cite{VanLoi:2023utt}.

A detailed study of the collider phenomenology and flavor-changing neutral currents induced by the flavor-dependent couplings of $Z'$ has been carried out in Ref.~\cite{VanLoi:2023utt}. In particular, constraints from dilepton searches at the LHC and low-energy flavor observables were shown to push the $Z'$ mass to the multi-TeV regime. In the present work, we therefore adopt these bounds and focus on the DM phenomenology of the model.

\section{\label{sec:DM}Residual discrete symmetry and scenarios for two-component dark matter}
The VEVs of the singlet scalars $\chi_{1,2}$ spontaneously break the $U(1)_X$ gauge symmetry. However, this breaking is not complete, and a residual discrete symmetry remains. A generic $U(1)_X$ transformation can be written as $R=e^{i\delta X}$, where $\delta$ is a continuous transformation parameter. Requiring the invariance of the scalar VEVs, $R\langle\chi_1\rangle=\langle\chi_1\rangle$ and $R\langle\chi_2\rangle=\langle\chi_2\rangle$, leads to the condition $\delta=k\pi/z$ with $k$ being an integer. As a result, the residual symmetry reduces to $R=(-1)^{kX/z}$. This symmetry is also preserved by the VEV of the SM Higgs doublet, since $R\langle H\rangle=\langle H\rangle$. The residual symmetry is further isomorphic to a discrete group, $Z_2=\{1,p_X\}$, with $p_X=(-1)^{X/z}$ and $p_X^2=1$. Taking into account the spin-parity group $S=\{1,p_s\}$, where $p_s=(-1)^{2s}$ is always conserved due to Lorentz invariance, the full symmetry structure can be written as $Z_2\otimes S\cong[(Z_2\otimes S)/\mathbb{Z}_2]\otimes\mathbb{Z}_2$. The physically relevant invariant discrete subgroup is 
\be \mathbb{Z}_2=\{1,p\}, \ee
with $p=(-1)^{X/z+2s}$ and $p^2=1$. Since the quotient group $(Z_2\otimes S)/\mathbb{Z}_2=\{\{1,p\},\{1,p_s\}\}$ remains conserved whenever $\mathbb{Z}_2$ is conserved, we identify this $\mathbb{Z}_2$ as the effective residual symmetry of the model. Under the residual $\mathbb{Z}_2$ symmetry, all SM fields, the right-handed neutrino $\nu_{3R}$, and the scalars $\chi_1$ and $\chi_2$ are even ($p = +1$), while the right-handed neutrinos $\nu_{1,2R}$ and the inert scalars $\phi$ and $\eta$ are odd ($p = -1$). The $\mathbb{Z}_2$ transformation properties of all particles are summarized in Table~\ref{parity}.
\begin{table}[h]
\bc
\begin{tabular}{c|ccccccccccccccc}
\hline\hline
Field &$\nu_{aL}$ & $e_a$ & $u_a$ & $d_a$ & $H^+$ & $H^0$ & $\nu_{1,2R}$ & $\nu_{3R}$ & $\chi_1$ & $\chi_2$ & $\ph^0$ & $\ph^-$ & $\eta$\\
\hline
$\mathbb{Z}_2$ &$+$ & $+$ & $+$ & $+$ & $+$ & $+$ & $-$ & $+$ & $+$ & $+$ & $-$ & $-$ & $-$\\
\hline\hline
\end{tabular}
\caption[]{\label{parity}Transformation properties of particles under the residual $\mathbb{Z}_2$ symmetry.}
\ec
\end{table}

Because $\mathbb{Z}_2$ is an exact residual symmetry of the theory, the lightest $\mathbb{Z}_2$-odd particle is stable and thus provides a natural DM candidate. In addition, owing to gauge invariance, the right-handed neutrino $\nu_{3R}$ has no allowed decay channels into SM fields and is therefore also stable, providing a second DM component. Consequently, the model naturally predicts two possible realizations of multicomponent DM:
\begin{itemize}
    \item a two-fermion DM scenario, in which both DM components are fermionic;
    \item a fermion--scalar DM scenario, in which one component is fermionic and the other is scalar.
\end{itemize}

\section{\label{sec:relic}Two-component dark matter phenomenology}
In this section, we investigate the phenomenology of two-component DM in the present model. Since the DM candidates interact appreciably with SM particles in the thermal bath of the early Universe through both gauge and Higgs portals, and since the new-physics scale is expected to lie in the TeV range--potentially accessible at current collider experiments--we focus on WIMP scenarios. In this framework, the relic abundance of each DM component is determined via the standard thermal freeze-out mechanism. 

We consider a two-component DM system consisting of a generic $\mathbb{Z}_2$-odd particle $X_a$ and the right-handed neutrino $\nu_{3R}$. The particle $X_a$ may correspond to any $\mathbb{Z}_2$-odd state, either a fermion $\nu_{1,2R}$ or a scalar ($R_{1,2},I_{1,2},H^{\pm}$). The evolution of their number densities is governed by a coupled set of Boltzmann equations, 
\bea 
\fr{dn_{X_a}}{dt}+3Hn_{X_a}&=&-\langle\sigma v\rangle_{X_aX_a^c\to X_0X_0}(n_{X_{a}}^2-\bar{n}_{X_{a}}^2)\crn
&&-\langle\sigma v\rangle_{X_aX_a^c\to\nu_{3R}\nu_{3R}^c}\left(n_{X_{a}}^2-n_{\nu_{3R}}^2\bar{n}^2_{X_a}/\bar{n}^2_{\nu_{3R}}\right)\Theta(m_{X_a}-m_{\nu_{3R}})\crn 
&&-\langle\sigma v\rangle_{X_aX_b^c\to X_0 X_0}\left(n_{X_a}n_{X_b}-\bar{n}_{X_a}\bar{n}_{X_b}\right) \crn 
&&+\langle\sigma v\rangle_{\nu_{3R}\nu_{3R}^c\to X_aX_a^c}\left(n_{\nu_{3R}}^2-n_{X_{a}}^2\bar{n}^2_{\nu_{3R}}/\bar{n}^2_{X_{a}}\right)\Theta(m_{\nu_{3R}}-m_{X_{a}})\crn && -\langle\sigma v\rangle_{X_{a}\nu_{3R}^c\to X_a\nu_{3R}^c}\left(n_{\nu_{X_a}}n_{\nu_{3R}}-n_{X_a}n_{\nu_{3R}}\bar{n}_{X_a}/\bar{n}_{\nu_{3R}}\right)\Theta(m_{X_a}-m_{\nu_{3R}})\crn
&& +\langle\sigma v\rangle_{\nu_{3R}X_{a}^c\to \nu_{3R}X_a^c}\left(n_{\nu_{3R}}n_{X_a}-n_{X_a}n_{\nu_{3R}}\bar{n}_{\nu_{3R}}/\bar{n}_{X_a}\right)\Theta(m_{\nu_{3R}}-m_{X_a}), \crn  
 \fr{dn_{\nu_{3R}}}{dt}+3Hn_{\nu_{3R}}&=& -\langle\sigma v\rangle_{\nu_{3R}\nu_{3R}^c\to X_0X_0}(n_{\nu_{3R}}^2-\bar{n}_{\nu_{3R}}^2)\crn 
 && -\langle\sigma v\rangle_{\nu_{3R}\nu_{3R}^c\to X_aX_a^c}\left(n_{\nu_{3R}}^2-\bar{n}_{X_a}^2\bar{n}_{\nu_{3R}}^2/\bar{n}_{X_a}^2\right)\Theta(m_{\nu_{3R}}-m_{X_a})\crn 
 &&+\langle\sigma v\rangle_{X_aX_a^c\to\nu_{3R}\nu_{3R}^c}\left(n_{X_{a}}^2-n_{\nu_{3R}}^2\bar{n}^2_{X_a}/\bar{n}^2_{\nu_{3R}}\right)\Theta(m_{X_a}-m_{\nu_{3R}})\crn 
 &&  -\langle\sigma v\rangle_{\nu_{3R}X_{a}^c\to \nu_{3R}X_a^c}\left(n_{\nu_{3R}}n_{X_a}-n_{X_a}n_{\nu_{3R}}\bar{n}_{\nu_{3R}}/\bar{n}_{X_a}\right)\Theta(m_{\nu_{3R}}-m_{X_a})\crn 
 && +\langle\sigma v\rangle_{X_{a}\nu_{3R}\to X_a\nu_{3R}}\left(n_{\nu_{3R}}n_{X_a}-n_{X_a}n_{\nu_{3R}}\bar{n}_{X_a}/\bar{n}_{\nu_{3R}}\right)\Theta(m_{X_a}-m_{\nu_{3R}}),  
\eea 
which include the following processes:
\begin{itemize}
\item Pair annihilation of each DM component into $\mathbb{Z}_2$-even particles,
\item Coannihilation between different $\mathbb{Z}_2$-odd particles,
\item DM conversion processes between the two DM components.
\end{itemize}
In these equations, the thermally averaged annihilation cross section times relative velocity is denoted by $\langle\sigma v\rangle$. The Heaviside step function $\Theta$ accounts for the mass hierarchy between the two DM components, ensuring that conversion processes are included only when kinematically allowed. The quantities $n_i$ represent the number densities of each DM component (with $i=X_a,\nu_{3R}$), while $\bar{n}_i$ denote their corresponding equilibrium number densities. The symbol $X_b$ refers to a $\mathbb{Z}_2$-odd particle different from $X_a$, and $X_0$ denotes $\mathbb{Z}_2$-even particles.

For the $\mathbb{Z}_2$-odd component $X_a$, the relic abundance is determined by all possible annihilation channels into even states, coannihilation with other odd particles, and conversion processes into $\nu_{3R}$. In contrast, the right-handed neutrino $\nu_{3R}$, being stabilized by gauge invariance rather than by the residual $\mathbb{Z}_2$ symmetry, participates only in annihilation and DM conversion processes, but does not undergo coannihilation with other $\mathbb{Z}_2$-odd states in the same way as $X_a$.

The coupled Boltzmann equations are solved numerically using the public package micrOMEGAs 6.2.4, which provides a consistent treatment of annihilation, coannihilation, and DM conversion effects in multi-component DM frameworks~\cite{Alguero:2023zol}. Before presenting the numerical results, we specify the assumptions adopted for the input parameters. The quartic couplings $\la_0, \la_1, \la_2$, $\la_3, \la_4,\la_5, \la_6$, and $\la_7$ do not play a significant role in the DM phenomenology considered here and are therefore omitted from the numerical analysis. In contrast to Ref.~\cite{VanLoi:2023utt}, we assume comparable singlet VEVs, $\Lambda_1 \sim \Lambda_2$, which motivates taking the couplings between the dark scalars $\phi, \eta$ and the singlet fields $\chi_1, \chi_2$ to be of the same order, namely $\la_{10}\sim \la_{12}$ and $\la_{11}\sim \la_{13}$. Similarly, we assume $\la_{8}\sim \la_{15}$, since both control interactions between the SM Higgs doublet $H$ and the dark scalar $\phi$. The scalar mixing angles $\epsilon_R$ and $\epsilon_I$ are taken to be small, $\epsilon_{R,I} \ll 1$, as a consequence of the hierarchy $v\ll \mu\sim \La_1, \La_2$ together with moderately small quartic couplings $\la \ll 1$, leading to an approximate alignment in which the dark doublet $\phi$ and singlet $\eta$ predominantly correspond to the mass eigenstates $R_1, I_1$ and $R_2, I_2$, respectively. In the fermionic DM sector involving $\nu_{1,2,3R}$, the Yukawa couplings $f^{\nu}_{\alpha\beta}$ are assumed to be flavor diagonal for simplicity. Under these assumptions, the parameters most relevant for the DM phenomenology are $\la_7,\la_{11},\la_{14}, \mu, \la $, the Yukawa couplings $h^{\nu}_{ab}$ and $f^{\nu}_{11,22,33}$, as well as the DM component masses $m_{\nu_{1,2,3R}}$ and $m_{I_2}$. 

The relevant scalar couplings are varied within the following ranges in order to satisfy the perturbativity condition:  
\bea
&&|\la_{8,9,10,11,14}|<4\pi, \hs \mu \in [1,20] \ \text{TeV}, \hs \la \in [10^{-4},10^{-1}], \hs h^{\nu}_{a\beta} \in [10^{-3},10^{-2}] , \crn
&& m_{I_2} \in [0.5,10] \ \text{TeV}, \hs m_{\nu_{3R}} \in [0.5,10] \ \text{TeV}, \hs \mu\sim \La_1,\La_2 \in [1,20] \ \text{TeV}.
\eea
We emphasize that the allowed ranges of $\la$ and $h^{\nu}_{a\beta}$ are more restricted than those of the other couplings, since they enter directly into the one-loop neutrino mass matrix in the scotogenic mechanism. In order to reproduce the observed tiny active neutrino masses without requiring excessive fine-tuning among parameters, these couplings must be moderately suppressed, as discussed explicitly in Ref.~\cite{VanLoi:2023utt}.
\subsection{Two-fermion dark matter}
We first consider the scenario in which both DM components are fermions. For simplicity, we assume that the right-handed neutrinos $\nu_{1R}$ and $\nu_{2R}$ do not mix. Without loss of generality, we take $\nu_{1R}$ to be the lightest $\mathbb{Z}_2$-odd particle. The DM sector is therefore composed of $\nu_{1R}$ and $\nu_{3R}$, realizing a purely fermionic two-component DM scenario.
\begin{figure}[H]
	\begin{center}
		\includegraphics[scale=0.3]{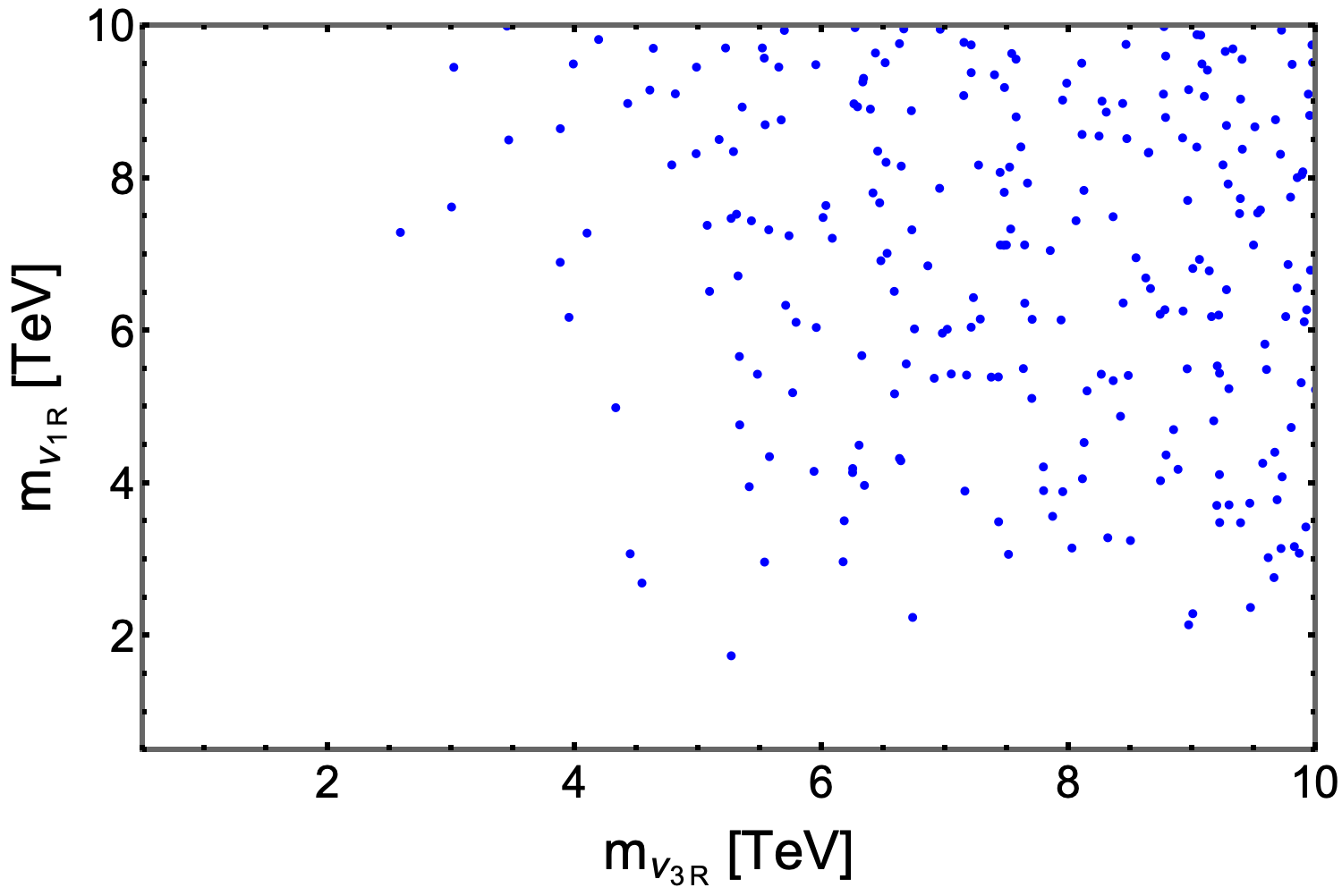}
		\caption{\label{fs1}Correlation between the masses of the two fermionic DM components, $m_{\nu_{1R}}$ and $m_{\nu_{3R}}$, that reproduce the observed total relic abundance, $\Omega_{\nu_{1R}}h^2 + \Omega_{\nu_{3R}}h^2 \simeq 0.12$~\cite{Planck:2018vyg}.}	
	\end{center}
\end{figure}

In Fig.~\ref{fs1}, we display the correlation between the two fermionic DM masses, $m_{\nu_{1R}}$ and $m_{\nu_{3R}}$, that reproduces the observed relic abundance, $\Omega_{\nu_{1R}} h^2 + \Omega_{\nu_{3R}} h^2 \simeq 0.12$~\cite{Planck:2018vyg}. The allowed parameter space is predominantly located in the upper-right region of the plane, which translates into approximate lower bounds of $m_{\nu_{1R}} \gtrsim 2$ TeV and $m_{\nu_{3R}} \gtrsim 2.5$ TeV.

We find that, over most of the viable parameter space, the two fermionic components contribute comparably to the total relic density. This feature can be traced back to the absence of a strong hierarchy in the underlying mass scales, $\La_1 \sim \La_2$, which leads to annihilation cross sections of similar magnitude for $\nu_{1R}$ and $\nu_{3R}$.
This behavior is qualitatively different from our previous study with $\La_2 \gg \La_1$, where a hierarchical mass choice, $m_{\nu_{3R}} = 6 m_{\nu_{1R}}$, was adopted. In that case, the relic fraction of $\nu_{1R}$ could vary widely, indicating a much more asymmetric contribution between the two fermionic components.

\begin{figure}[H]
	\begin{center}
		\includegraphics[scale=0.3]{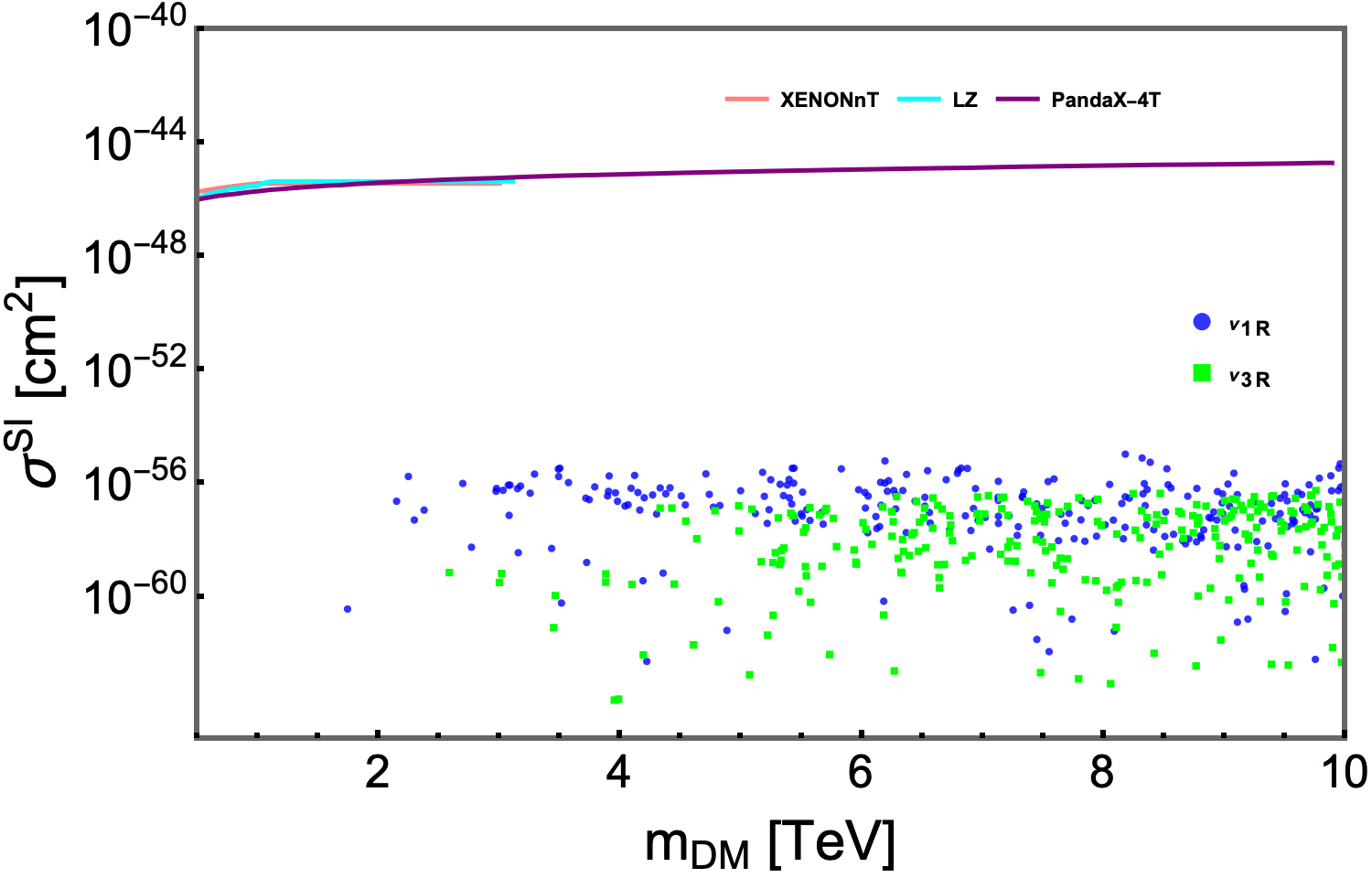}
		\caption{\label{fs2}Spin-independent scattering cross sections of the two fermionic DM components, $\nu_{1R}$ and $\nu_{3R}$, as functions of their masses. The colored solid lines represent the current upper limits from direct-detection experiments, namely XENONnT~\cite{XENON:2025vwd}, LZ~\cite{LZ:2022lsv}, and PandaX-4T~\cite{PandaX:2024qfu}.}	
	\end{center}
\end{figure}

In addition to relic density constraints, direct-detection experiments, which measure the scattering cross section of DM particles off nucleons in target nuclei, provide important bounds on DM properties. In Fig.~\ref{fs2}, we present the spin-independent (SI) scattering cross sections $\sigma^{\text{SI}}$ for the two fermionic DM components, $\nu_{1R}$ (blue points) and $\nu_{3R}$ (green points), as functions of their masses. We find that the predicted SI cross sections lie in the range $\sigma^{\text{SI}}_{\nu_{1R},\nu_{3R}} \sim \mathcal{O}(10^{-56} - 10^{-60})$ cm$^2$, which are several orders of magnitude below the current upper limits reported by the XENONnT~\cite{XENON:2025vwd}, LZ~\cite{LZ:2022lsv}, and PandaX-4T~\cite{PandaX:2024qfu} collaborations.

This strong suppression can be understood as follows. The dominant contributions to the elastic scattering between the fermionic DM candidates and nucleons arise from $t$-channel exchange of the heavy scalar states $H_1$ and $H_2$. The corresponding effective interactions depend on the couplings $\bar{q} q H_{1,2}$ ($q = u, d$), which are proportional to the light quark masses and therefore suppressed by factors of order $\mathcal{O}(m_q/M) \ll 1$, where $M$ denotes the new-physics scale. As a result, the resulting SI cross sections are naturally tiny.

It is worth emphasizing that, in the present work, we include the additional contributions from the heavy scalar mediators $H_1$ and $H_2$ to the DM–nucleon scattering amplitude. In contrast, the previous study in Ref.~\cite{VanLoi:2023utt} considered only the $Z'$ portal, which leads predominantly to spin-dependent (SD) interactions. Moreover, due to the vector-like structure of the $Z'$ couplings to quarks in that setup, the corresponding direct-detection signals were strongly suppressed.

In conclusion, the predicted spin-independent cross sections for the two fermionic DM components are well below current experimental sensitivities and therefore fully consistent with existing direct-detection constraints.

\subsection{Fermion--scalar dark matter}
We next turn to the fermion--scalar DM scenario, in which one DM component is fermionic while the other is scalar. This situation arises when the lightest $\mathbb{Z}_2$-odd particle is a scalar, identified as $I_2$, whereas the accidentally stable right-handed neutrino $\nu_{3R}$ constitutes the fermionic component. We stress that this possibility was not explored in our previous work, where the hierarchical limit $\La_2 \gg \La_1$ rendered all $\mathbb{Z}_2$-odd scalars heavy and led exclusively to a purely fermionic two-component DM setup. In the present framework with $\La_1 \sim \La_2$, the scalar sector becomes phenomenologically relevant, allowing for a genuinely new fermion--scalar two-component DM realization.

From the scalar mass spectrum given in Eqs.~(\ref{mri1}--\ref{mhc}), one can see that $m_{I_2}$ is always the lightest among the $\mathbb{Z}_2$-odd scalar states. In particular, its squared mass differs from the other scalar masses by a term proportional to $\sqrt{2}\mu \Lambda_2$. Since $\mu \sim \Lambda_1 \sim \Lambda_2$ and these scales lie at the TeV level, this contribution is sizable and ensures that $I_2$ remains the lightest dark scalar. Being $\mathbb{Z}_2$-odd, $I_2$ is therefore stable and naturally identified as the scalar DM component. The resulting fermion--scalar DM system thus consists of $\nu_{3R}$ and $I_2$.

The presence of a scalar DM particle significantly enriches the dynamics of the system. In addition to the standard annihilation channels of $\nu_{3R}$, the scalar $I_2$ can annihilate efficiently through Higgs-portal interactions and can participate in conversion processes such as $\nu_{3R}\nu_{3R} \leftrightarrow I_2 I_2$. These additional annihilation and conversion channels lead to a coupled freeze-out pattern that differs qualitatively from the purely fermionic case, where the available interactions are more restricted.

\begin{figure}[H]
	\begin{center}
		\includegraphics[scale=0.3]{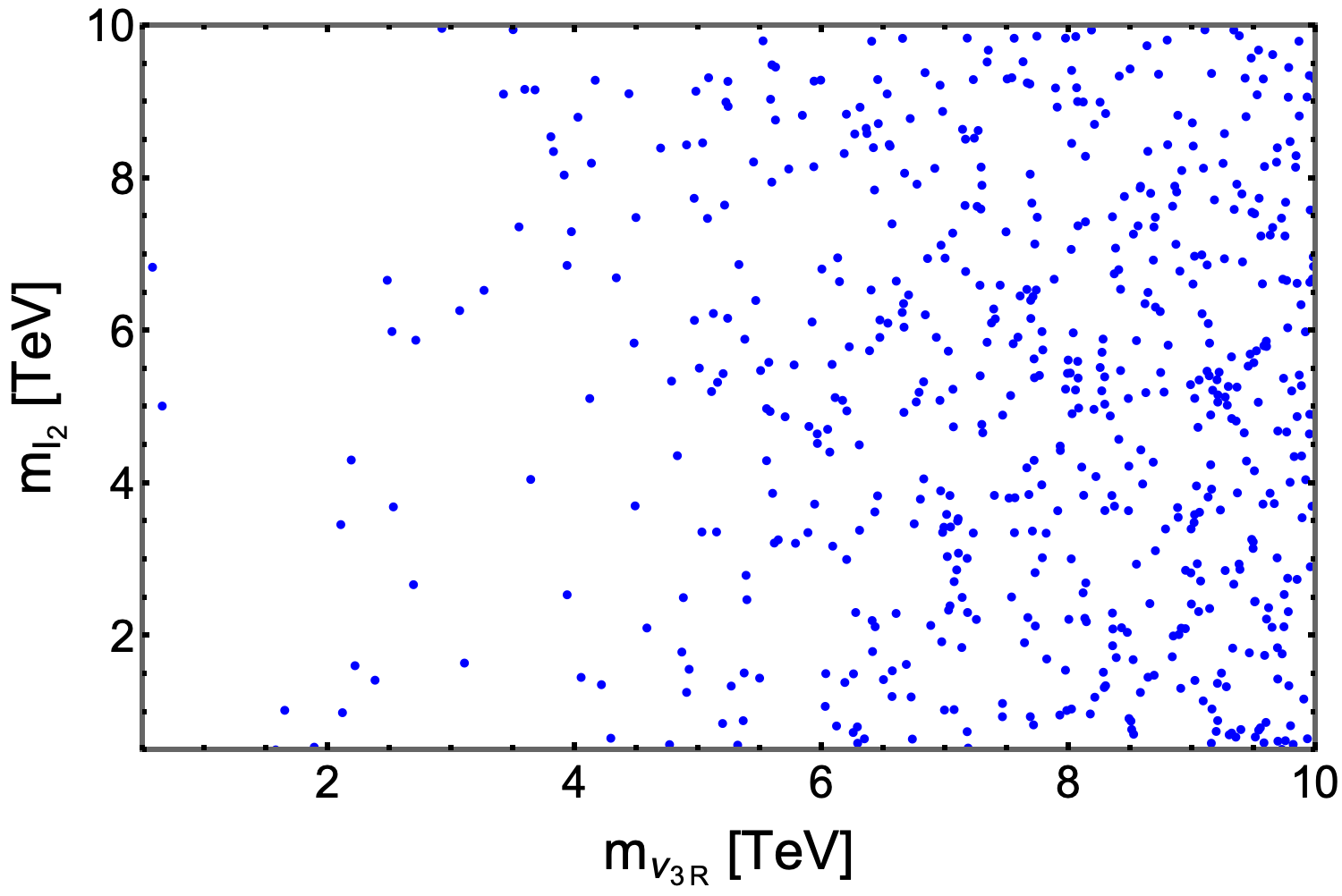}
		\caption{\label{fs3}Correlation between the masses of the fermionic DM component $m_{\nu_{3R}}$ and the scalar component $m_{I_2}$ that reproduce the observed total relic abundance, $\Omega_{\nu_{3R}} h^2 + \Omega_{I_2} h^2 \simeq 0.12$~\cite{Planck:2018vyg}.}	
	\end{center}
\end{figure}

In Fig.~\ref{fs3}, we present the allowed parameter space in the $(m_{\nu_{3R}}, m_{I_2})$ plane that reproduces the observed total relic abundance, $\Omega_{\nu_{3R}} h^2 + \Omega_{I_2} h^2 \simeq 0.12$~\cite{Planck:2018vyg}. We find that the fermionic component typically requires $m_{\nu_{3R}} \gtrsim 1.65$ TeV, while the scalar mass $m_{I_2}$ can span a broad range within the scanned parameter space. The comparatively weaker lower bound on $m_{\nu_{3R}}$, relative to the two-fermion case, can be attributed to the additional scalar-mediated annihilation and conversion channels, which help deplete the total DM abundance more efficiently.

\begin{figure}[H]
	\begin{center}
		\includegraphics[scale=0.3]{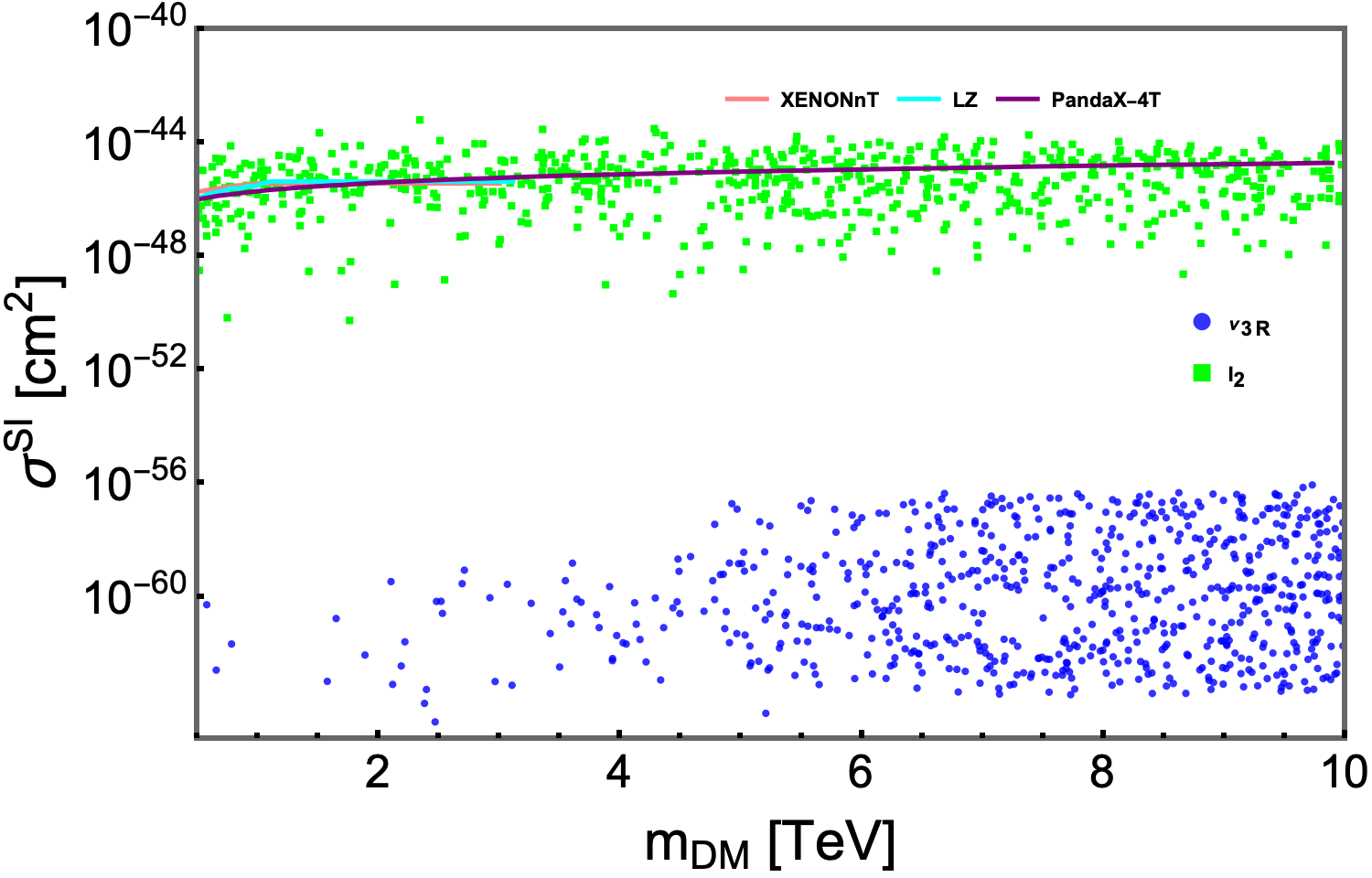}
		\caption{\label{fs4} Spin-independent cross sections of the fermionic DM component $\nu_{3R}$ and the scalar component $I_2$ as functions of their masses. The colored solid lines represent the current upper limits from direct-detection experiments, namely XENONnT~\cite{XENON:2025vwd}, LZ~\cite{LZ:2022lsv}, and PandaX-4T~\cite{PandaX:2024qfu}.}	
	\end{center}
\end{figure}

Turning to the constraints from direct-detection experiments, the results are displayed in Fig.~\ref{fs4}. As in the purely fermionic case, the SI cross section of the fermionic component $\nu_{3R}$ remains well below the current experimental limits and is therefore phenomenologically safe.

The situation is qualitatively different for the scalar DM component $I_2$. Its SI cross section is significantly larger than that of $\nu_{3R}$ and, in certain regions of parameter space, approaches or even exceeds the present bounds. The scattering of $I_2$ off nucleons proceeds dominantly via $t$-channel exchange of the SM-like Higgs boson $h$ and the heavy Higgs bosons $H_1$ and $H_2$. The corresponding amplitudes are mainly controlled by the scalar couplings $\la_9$ and $\la_{11}$. Importantly, these same couplings also govern the dominant annihilation channels of $I_2$,
\bea
I_2 I_2 \to hh, W^+W^-, ZZ, t\bar{t},
\eea
which determine its relic abundance. This leads to a non-trivial interplay between relic density and direct-detection constraints. Sufficiently large values of $\la_9$ and $\la_{11}$ enhance the annihilation cross section and prevent $\Omega_{I_2} h^2$ from overclosing the Universe. On the other hand, the same enhancement increases the SI scattering rate, potentially pushing $\sigma^{\text{SI}}_{I_2}$ above the experimental limits.

This tension significantly restricts the viable parameter space of the fermion–scalar scenario and makes it considerably more constrained than the purely fermionic case. Although the current data still allow consistent solutions, the predicted SI cross section for $I_2$ typically lies close to the projected sensitivity of next-generation direct-detection experiments, $\sigma^{\text{SI}} \lesssim \mathcal{O}(10^{-47} - 10^{-48})$. Therefore, this new realization of two-component DM can be decisively tested--and potentially ruled out--in the near future.

\section{\label{sec:conclusion}Conclusions}

We have studied the DM phenomenology of a flavor-dependent $U(1)_X$ extension of the SM in which anomaly cancellation predicts three fermion generations and requires three right-handed neutrinos. In contrast to Ref.~\cite{VanLoi:2023utt}, where a strong hierarchy $\Lambda_2 \gg \Lambda_1$ led exclusively to a purely fermionic two-component DM scenario, we have considered here a more general scalar spectrum with $\Lambda_1 \sim \Lambda_2$. This relaxation gives rise to a new realization of two-component DM consisting of one fermionic component ($\nu_{3R}$) and one scalar component ($I_2$), in addition to the purely fermionic case.

By solving the coupled Boltzmann equations numerically, we find that the two-fermion scenario is compatible with the observed relic abundance for TeV-scale masses and predicts highly suppressed spin-independent cross sections, well below current experimental limits. In contrast, the fermion--scalar scenario exhibits a non-trivial interplay between relic-density and direct-detection constraints, since the same scalar couplings control both annihilation and nucleon-scattering processes. While still consistent with present data, this scenario can be decisively tested by next-generation direct-detection experiments. Overall, relaxing the scalar-sector hierarchy significantly enriches the DM structure of the model and enhances its experimental testability.

\section*{Acknowledgements}
This research is funded by Phenikaa University under grant number PU2024-4-A-03.

\bibliography{combine}

\end{document}